\documentstyle[aps]{revtex}
\input{epsf.tex}

\def\be{\begin{equation}}
\def\ee{\end{equation}}
\def\ba{\begin{array}}
\def\ea{\end{array}}
\def\bea{\begin{eqnarray}}
\def\eea{\end{eqnarray}}
\begin{document}
\draft
\title{\bf Cluster-decay of hot $^{56}Ni^*$ formed in $^{32}S+^{24}Mg$ reaction}
\author{Raj K. Gupta$^{a,b,c}$, Rajesh Kumar$^{a}$, Narinder K. Dhiman$^{a}$,
M. Balasubramaniam$^{a,b}$, Werner Scheid$^b$ and C. Beck$^c$\\
{\it $^a$ Physics Department, Panjab University, Chandigarh-160014,}
{\it India.}\\
{\it $^b$ Institut f\"ur Theoretische Physik, Justus-Liebig-Universit\"at,}\\
{\it Heinrich-Buff-Ring 16, D-35392, Giessen, Germany.}\\
{\it $^c$ Institut de Recherches Subatomiques, UMR7500, IN2P3/ Universit\'e}\\
{\it Louis Pasteur, F-67037 Strasbourg, France.}\\
}
\date{\today}
\maketitle

\begin{abstract}
The decay of $^{56}Ni^*$, formed in $^{32}S+^{24}Mg$ reaction at the incident
energies $E_{cm}$=51.6 and 60.5 MeV, is calculated as a cluster decay process
within the Preformed Cluster-decay Model (PCM) of Gupta et al. re-formulated
for hot compound systems. Interesting enough, the cluster decay process is
shown to contain the complete structure of both the measured fragment cross
sections and total kinetic energies (TKEs). The observed deformed shapes of
the exit channel fragments are simulated by introducing the neck-length
parameter at the scission configuration, which nearly coincides the $^{56}Ni$
saddle configuration. This is the only parameter of the model, which though
is also defined in terms of the binding energy of the hot compound system and 
the ground-state binding energies of the various emitted fragments. 
For the temperature effects included in shell corrections only, the normalized 
$\alpha$-nucleus 
s-wave
cross sections calculated for nuclear shapes
with outgoing fragments separated within nuclear proximity limit (here
$\sim$0.3 fm) can be compared with the experimental data, and the TKEs are found 
to be in reasonably good agreement with experiments for the angular momentum 
effects added in the sticking limit for the moment of inertia. The incident 
energy effects are also shown in predicting different separation distances and
angular momentum values for the best fit. Also, some light particle production
(other than the evaporation residue, not treated here) is predicted at these
energies and, interestingly, $^4He$, which belongs to evaporation residue, is 
found missing as a dynamical cluster-decay fragment. Similar results are 
obtained for temperature effects included in all the terms of the potential 
energy. The non-$\alpha$ fragments are now equally important and hence present 
a more realistic situation with respect to experiments.
\end{abstract}
\par\noindent
PACS Nos. {25.70.Jj, 23.70.+j, 24.10.-i, 23.60.+e} 

\section{INTRODUCTION}
Experimentally, $^{56}Ni$ is an extensively studied compound system by using
different entrance channels, namely $^{16}O+^{40}Ca$, $^{28}Si+^{28}Si$ and
$^{32}S+^{24}Mg$, and at various incident energies ranging from 1.5 to 2.2 
times the Coulomb barrier (see, e.g., the review \cite{sanders99} and the 
other direct and more recent Refs. 
\cite{sanders87,sanders89,sanders94,nouicer99,beck01,thummerer01,bhatta01}).
At such incident energies, the incident flux is found to get trapped by the
formation of a compound nucleus (CN), which is in addition to a significant
large-angle elastic scattering cross section. For lighter masses ($A_{CN}<44$),
such a compound nucleus decays subsequently by the emission of mainly light
particles (n, p, $\alpha$) and $\gamma$-rays; i.e. with very small component
of heavy fragment ($A>4$) emission. An experimental measure of this so-called
particle evaporation residue yield is the CN fusion cross section.  For
somewhat heavier systems, like $^{48}Cr$ and $^{56}Ni$, a significant decay
strength to $A>4$ fragments, the mass-asymmetric channels, is also observed
which could apparently not arise from a direct reaction mechanism because of
the large mass-asymmetry differences between the entrance and exit channels.
The measured angular distributions and energy spectra are consistent with
fission-like decays of the respective compound systems.

For the $^{32}S+^{24}Mg\rightarrow ^{56}Ni^*$ reaction, in one of the
experiments, the mass spectra for A=12 to 28 fragments and the total kinetic
energy (TKE) for only the most favoured (enhanced yields) $\alpha$-nucleus
fragments are measured at the energies $E_{lab}=$ 121.1 and 141.8 MeV, or
equivalently at $E_{cm}=$51.6 and 60.5 MeV, respectively \cite{sanders87,sanders89}.
Note that $^{56}Ni$ is a negative Q-value system (negative $Q_{out}$,
different for different exit channels) and hence would decay only if it were
produced in heavy ion reactions with sufficient compound nucleus excitation
energy $E_{CN}^*$ ($=E_{cm}+Q_{in}$), to compensate for the negative $Q_{out}$,
the deformation energy of fragments $E_d$, their total kinetic energy TKE and
the total excitation energy TXE, in the exit channel, as
\be
E_{CN}^*=\mid Q_{out}(T)\mid+E_d(T)+TKE(T)+TXE(T);
\label{eq:1}
\ee
see Fig.1 where $E_d$ is neglected because fragments are considered to be
spherical. Here $Q_{in}$ is positive (=16.68 MeV for $^{32}S+^{24}Mg$ 
entrance channel) and hence adds to the entrance channel kinetic energy 
$E_{cm}$ of the two incoming nuclei in their ground states. In another 
experiment \cite{sanders94} for $^{32}S+^{24}Mg$ reaction at $E_{cm}=$51.0 
and 54.5 MeV, the excitation-energy spectra for only the symmetric 
$^{28}Si+^{28}Si$ and near-symmetric $^{24}Mg+^{32}S$ channels are measured, 
whose analysis indicate that a specific set of states in $^{28}Si$ correspond 
to highly deformed bands. In other words, the expected shapes of some of the 
observed fragments in the exit channel could be relatively deformed. It is 
interesting to note that this result is supported by a very recent study of 
the $^{28}Si+^{28}Si$ reaction at $E_{cm}=$ 55 MeV, where the population of 
highly excited states in the $^{24}Mg$, $^{28}Si$ and $^{32}S$ nuclei 
indicated a selective and enhanced population of deformed bands \cite{beck01}.
In a still other recent experiment \cite{thummerer01}, the incident energy
used in the same $^{32}S+^{24}Mg\rightarrow ^{56}Ni^*$ reaction is $E_{lab}=$
130 MeV and an enhanced emission yield by a factor of 1.5 to 1.8 is observed
for $^8Be$ over the two $\alpha$-particles. The aim of our present work is to
understand some of the results of these experiments.

The above stated light particles ($A\le 4$) production, the
evaporation residue, is very satisfactorily understood as the equliberated
compound nucleus emission in the statistical Hauser Feshbach analysis
\cite{sanders89,charity88,campo91,sanders91,matsuse97}, using the LILITA or
CASCADE codes. The Hauser Feshbach calculations are also extended to include
the complex fragments, like the ones observed in the experiments mentioned
above. These are considered in the, so-called, BUSCO code \cite{campo91} or
the Extended Hauser-Feshbach scission-point model \cite{matsuse97}. Within the
framework of the Extended Hauser-Feshbach method \cite{matsuse97}, the above
noted observed enhanced emission of $^8Be$ over the evaporation of two
$\alpha$-particles in $^{32}S+^{24}Mg$ reaction is shown related to an
increased deformation of the heavier fragment $^{48}Cr$ \cite{thummerer01}.
The emission of complex fragments ($A>4$, also called the intermediate mass 
fragments, IMFs, or "clusters") is alternatively treated as the binary fission of a
compound nucleus in the statistical fission models \cite{moretto75,vandenbosch73},
using the GEMINI code \cite{charity88} or the saddle-point "transition-state"
model \cite{sanders89,nouicer99,sanders91}. The transition-state model,
treating the complex fragments emission as a compound-nucleus fission process
(the fusion-fission) seems to explain the observed mass spectra and
excitation-energy spectra rather well for the $^{32}S+^{24}Mg$ reaction at
the two energies used in respective experiments \cite{sanders89,sanders94}.
Also, the measured TKE for the symmetric fission is comparable to the
saddle-point potential energy at $\ell =36\hbar$ \cite{sanders89}. Then,
there are other processes, like the deep-inelastic (DI) orbiting or scattering,
that have also been studied for this reaction but do not seem to explain the
observed data \cite{sanders89}.

In the statistical fission models \cite{moretto75,vandenbosch73}, the fission
decay of a compound nucleus is determined by the phase space (level density)
available at the "transition" configuration, which is saddle or scission in
these models. For light systems, this choice can lead to a significant
population of many energetically allowed mass channels, though there is no
structure information of the compound system in these fission models. However,
the structure effects of the compound system seem to influence the observed
yields strongly since strong resonance behaviour is observed in the measured
excitation functions of large-angle elastic and inelastic scattering yields
in several light systems (see, e.g., \cite{beck01}). One possibility to account
for such structure effects is via the process of fragments (or clusters)
preformation in a compound nucleus and its subsequent decay as a cluster decay
process, proposed recently by some of us \cite{sharma00,gupta02}. The structure
information enters the process via the preformation probability (also, known as
the spectroscopic factors) of the fragments. We follow this approach of
preformed cluster decay \cite{sharma00,gupta02} here in this paper.

The cluster decay process was recently studied \cite{sharma00} for the
compound system $^{56}Ni^*$, using the preformed cluster-decay model (PCM) of 
Gupta and collaborators \cite{gupta88,malik89,gupta91,kumar94,gupta99a}. It was
shown that for the decay of $^{56}Ni^*$, the two processes of binary fission
(the dynamical collective mass transfer calculated, by some of us
\cite{gupta84,saroha85,malik86,puri92}, in the quantum mechanical
fragmentation theory \cite{maruhn74,gupta75,gupta99b}) and cluster decay are
almost indistinguishable, particularly at higher angular momenta. However,
this work was a simple model study where the role of TKE was analysed and
found to be significant for $\alpha$-nucleus structure in the measured yields.
This model is more recently re-formulated \cite{gupta02} for the IMFs emitted
from an excited $^{116}Ba^*$ compound nucleus produced in low energy
$^{58}Ni+^{58}Ni$ reaction. The IMFs in $^{116}Ba^*$ are shown to be produced
as multiple "clusters" of masses $A<20$ and only at $E_{lab}>200$ MeV, in 
agreement with experiments. Both of these works \cite{sharma00,gupta02} show 
that the IMFs in decay of excited $^{116}Ba^*$ or the complete mass spectra 
in decay of excited $^{56}Ni^*$ have their origin in the macroscopic liquid 
drop energy (the shell effects are almost zero at the excitation energies 
involved). For $^{116}Ba^*$ decay, the light particles ($Z\le 2$) emission, 
other than the promptly emitted via the statistical evaporation
process (not treated in this model), is also shown to be given, but at higher
energies where only the pure liquid drop energies enter the calculations.
Thus, the macroscopic liquid drop energy ($V_{LDM}$) is shown playing the most
important role in the cluster decay calculations. Apparently, the
compound nucleus being hot at the energies involved, the $V_{LDM}$ should also
depend on the temperature T. This is done here in this paper for the decay of
$^{56}Ni^*$ formed in $^{32}S+^{24}Mg$ reaction at the two energies,
$E_{cm}=$51.6 and 60.5 MeV \cite{sanders87,sanders89}. Also, the other terms of the
potential, that constitute the scattering potential $V(R)$, are considered
T-dependent. 

The T-dependent liquid drop model used is that of Davidson et al.
\cite{davidson94} which is based on the semi-empirical mass formula of Seeger
\cite{seeger61}. The model parameters of Seeger's formula at T=0 are re-fitted 
in view of the present availability of a larger data set for binding energies
\cite{audi95}. For the T-dependence in $V(R)$, we follow Davidson et al.
\cite{davidson94} and some other authors \cite{royer92}, discussed below.
The deformation effects of the fragments (and the neck formation between them)
are included here within the extended model of Gupta and collaborators
\cite{khosla90,gupta97,kumar97}, via a neck-length parameter at the scission
configuration which simulates the two centre nuclear shape parametrization,
used for both the light and heavy nuclear systems. A similar method has been
used earlier by other authors \cite{sanders89,sanders91,matsuse97}, discussed
below.

The dynamical cluster decay model for hot compound systems, a re-formulation
of the preformed cluster-decay model (PCM) of Gupta and co-workers
\cite{gupta88,malik89,gupta91,kumar94,gupta99a} for ground-state decays, is
presented in section 2 and its application to the hot $^{56}Ni^*$ nucleus data
from Refs. \cite{sanders87,sanders89}
in section 3. The (statistical) evaporation of light particles, that occur
promptly before the beginning of the binary decay process of cluster emission
studied here, is not included in this paper. Hence, any discussion of light
particles emission is that of one which is in addition to the ones emitted
promptly. Finally, a summary of our results is presented in section 4.

\section{THE DYNAMICAL CLUSTER DECAY MODEL FOR HOT COMPOUND SYSTEMS}
The cluster decay model developed here is the preformed cluster-decay model 
(PCM) of Gupta et al. \cite{gupta88,malik89,gupta91,kumar94,gupta99a} for 
the ground-state decays, re-formulated for hot and excited compound systems.
In this model, we treat the complex fragments (the IMFs or clusters) as
dynamical collective mass motion of preformed fragments through the barrier.
It is based on the well known dynamical (or quantum mechanical) fragmentation
theory \cite{maruhn74,gupta75,gupta99b} developed for fission and heavy ion
reactions, and used later for predicting the exotic cluster radioactivity
\cite{sandu80,rose84,gupta94} also. This theory is worked out in terms of the 
collective coordinates of mass asymmetry $\eta={{(A_1-A_2)}/{(A_1+A_2)}}$ and 
relative separation R, which in a PCM allows to define the decay half-life 
$T_{1\over 2}$, or the decay constant $\lambda $, as
\be
\lambda ={{{ln 2}\over {T_{1\over 2}}}}=P_0\nu _0 P, 
\label{eq:2}
\ee
where $P_0$, the preformation probability, refers to $\eta$-motion and P, 
the penetrability, to R-motion. Apparentlty, the two motions are taken as
decoupled, an assumption justified in our earlier works
\cite{maruhn74,gupta75,saroha86}. The $\nu _0$ is the barrier assault
frequency. 
In terms of the partial waves, the decay cross section 
\be
\sigma={\pi \over k^2}\sum_{\ell=0}^{\ell_{c}}(2\ell+1)P_0P; 
\qquad 
k=\sqrt{2\mu E_{c.m.}\over {\hbar^2}}
\label{eq:2a}
\ee
with $\mu =[A_1A_2/(A_1+A_2)]m={1\over 4}Am(1-\eta ^2)$ as the reduced mass 
and $\ell_{c}$, the critical (maximum) angular momentum, defined later. m is 
the nucleon mass. This means that $\lambda$ in (\ref{eq:2}) gives the s-wave
cross section, with a normalization constant $\nu _0$, instead of the 
${\pi}/{k^2}$ in (\ref{eq:2a}). However, in the present calculations, 
made for $\ell =0$ case, the normalization constant is obtained empirically
from the experimental data.

For $\eta$-motion, we solve the stationary Schr\"odinger equation in $\eta$,
at a fixed R,
\be
\{ -{{\hbar^2}\over {2\sqrt B_{\eta \eta}}}{\partial \over {\partial
\eta}}{1\over {\sqrt B_{\eta \eta}}}{\partial\over {\partial \eta
}}+V_R(\eta ,T)\} \psi ^{\nu}(\eta ) = E^{\nu} \psi ^{\nu}(\eta ),
\label{eq:3}
\ee
with $\nu$=0,1,2,3... and $R=R_a=C_t(=C_1+C_2)$, the first turning point,
fixed empirically for the ground-state (T=0) decay since this value of R
(instead of the compound nucleus radius $R_0$) assimilates to a good extent
the effects of both the deformations $\beta_i$ of two fragments and neck
formation between them \cite{kumar97}. In other words, the deformation
effects of the two fragments are included here in the scattering potential
V(R,T=0) for each $\eta$ by raising the first turning point $R_a$ from
$R_a=R_0$ to $R_a=C_t$ or $C_t+\sum\delta R(\beta_i)$, which is equivalent of 
lowering of the barrier, as is found to be the case for deformed fragments 
\cite{kumar97}. This method of inclusion of fragment deformation and the 
parametrization of the neck zone via a neck-length parameter $\delta R$ in the 
present calculations is quite similar to what has been achieved in both the
transition-state model of Sanders \cite{sanders89,sanders91} (in saddle point
configuration) and the Extended Hauser-Feshbach Method of Matsuse and
collaborators \cite{matsuse97} (in scission point configuration). It is also
shown in \cite{kumar97} that the alternative of calculating the fragmentation
potential $V(\eta)$ and scattering potential V(R) for deformed nuclei is not
practical since the experimental deformation parameters for all the possible
fragments ($A_1$,$A_2$), required for calculating $V(\eta)$, are generally not
available. The deformation effects of nuclei in our calculations are further
included via the S\"ussmann central radii $C_i=R_i-({b/R_i})$, with the radii
$R_i=1.28A_i^{1/3}-0.76+0.8A_i^{-1/3} fm$ and surface thickness parameter
$b=$0.99 fm. Note that the $C_t$ are different for different $\eta$-values
and hence $C_t$ is $C_t(\eta)$.

The eigen-solutions of Eq. (\ref{eq:3}) give the preformation probability 
\be
P_0={\sqrt {B_{\eta \eta}}}\mid \psi (\eta (A_i))\mid ^2\left ({2/A}\right ),
\label{eq:4}
\ee
(i=1 or 2), where $\psi (\eta )$ is $\psi ^{\nu=0}(\eta )$ if the ground-state
solution is chosen. However, the decay of $^{56}Ni$ in the ground-state
(T=0, $R_a=C_t$) is not allowed since $Q_{out}(T=0)$ is negative.

For the decay of a hot compound nucleus, we use an ansatz \cite{gupta02} for
the first turning point,
\be
R_a=C_t(\eta,T)+\Delta R(\eta,T),
\label{eq:5}
\ee
which depends on the total kinetic energy TKE(T). The corresponding potential 
$V(R_a)$ acts like an effective, positive Q-value, $Q_{eff}$, for the decay of 
the hot compound system at temperature T to two fragments in the exit channel 
observed in the ground-states (T=0). Thus, in terms of the respective binding 
energies B, $Q_{eff}$ is defined as
\bea
Q_{eff}(T)&=&B(T)-[B_1(T=0)+B_2(T=0)] \nonumber \\
          &=&TKE(T) \nonumber \\
          &=&V(R_a).
\label{eq:5a}
\eea
Since, $R_a=C_t(\eta)$ for T=0, $\Delta R(\eta)$ corresponds to the change in
TKE at $T$ with respect to its value at $T=0$ and hence can be estimated
exactly for the temperature effects included in the scattering potential V(R).
Note that in Eq. (\ref{eq:5}) $C_t$ is also taken to depend on temperature, as
is defined later in the following. Also, $\Delta R$ depends on $\eta$. In the
following, however, based on our earlier work \cite{gupta02}, instead, we use
a constant average value $\overline{\Delta R}$, independent of $\eta$, which
also takes care of the additional $\sum\delta R(\beta_i)$ effects of the 
deformations of fragments and neck formation between them. Note that 
$\overline{\Delta R}$ is the only parameter of the model, though it is shown 
that the structure of the calculated mass spectrum is nearly independent of the 
exact choice of this parameter value. The corresponding $Q_{eff}$ is denoted as 
$Q_{eff}(\overline{\Delta R})$.

In the above definition of $Q_{eff}$, apparently the two fragments would come
out of the barrier and go to ground state ($T\rightarrow 0$) only by emitting
some light particle(s) and/or $\gamma$-rays of energy, defined as (see Fig. 8)
\bea
E_x&=&B(T)-B(0) \nonumber\\
   &=&Q_{out}(T)-Q_{out}(T=0)+\Delta B \nonumber\\
   &=&Q_{eff}(T)-Q_{out}(T=0) \nonumber\\
   &=&TKE(T)-TKE(T=0).
\label{eq:12a}
\eea
Eq. (\ref{eq:12a}) means that one can also write
\bea
Q_{eff}(T)&=&TKE(T) \nonumber\\
          &=&Q_{out}(T=0)+E_x \nonumber\\
          &=&TKE(T=0)+E_x,
\label{eq:12b}
\eea
which is what one observes experimentally i.e. the fragments in the ground state 
with $Q_{out}(T=0)$ (=$TKE(T=0)$) and light particle(s) and $\gamma$-rays of 
energy $E_x$. The remaining excitation energy of the decaying system is then,
\be
E_{CN}^*-E_x=\mid Q_{out}(T)\mid+TKE(T=0)+TXE(T),
\label{eq:12c}
\ee
which again shows that the exit channel fragments are obtained with their TKE 
in the ground-state, i.e. with TKE(T=0). The excitation energy TXE(T) 
in (\ref{eq:12c}) is used in the secondary emission of light particles from 
the fragments, which are not treated here. Instead, we compare our 
calculations with the primary pre-secondary-evaporation fragments 
emission data. 

We notice from Eq. (\ref{eq:5a}) that for the ground-state (T=0) decay,
\be
Q_{eff}(T=0)=Q_{out}(T=0)=TKE(T=0),
\label{eq:5b}
\ee
as is the case for exotic cluster radioactivity \cite{gupta99a,gupta94}. In
fact, one can write Eq. (\ref{eq:5a}) as
\be
Q_{eff}(T)=Q_{out}(T)+\Delta B,
\label{eq:20}
\ee
where
\be
\Delta B=[B_1(T)+B_2(T)]-[B_1(T=0)+B_2(T=0)],
\label{eq:19}
\ee
the difference of binding energies at temperature T and the ground-state
binding energies of the two fragments. Also, for the ground-state (T=0)
decays, according to Eq. (\ref{eq:12a}), $E_x=0$ (no particle or $\gamma$-ray
emission), as is known to be true for exotic cluster radioactivity
\cite{gupta99a,gupta94}.

Thus, at temperature T, the preformation factor $P_0$ in Eq. (\ref{eq:4}) is 
calculated at $R_a=C_t(\eta)+\overline{\Delta R}$, with the temperature 
effects also included in $\psi (\eta )$ through a Boltzmann-like function
\be
\mid \psi \mid ^2 = \sum _{\nu =0}^{\infty}\mid\psi ^{\nu}\mid ^2
exp (-E^{\nu}/T),
\label{eq:6}
\ee
with the compound nucleus temperature $T$ (in MeV) related as
\be
E_{CN}^*=\left ({A/9}\right ){T}^2-T;
\label{eq:7}
\ee
and for the penetrability P, Eqs. (\ref{eq:5}) and (\ref{eq:5a}) for each
$\eta$ and T-values, mean that
\be
V(R_a)=V(C_t+\overline{\Delta R})=V(R_b)=Q_{eff}(\overline{\Delta R})=TKE(T),
\label{eq:9}
\ee
with $R_b$ as the second turning point, and penetrability $P$ calculated as
the WKB tunnelling probability for the path shown in Fig. 1 (or Fig. 8), as
\be
P=exp[-{2\over \hbar}{{\int }_{R_a}^{R_b}\{ 2\mu [V(R)-Q_{eff}]\}
^{1/2} dR}],
\label{eq:8}
\ee
solved analytically \cite{malik89}. 

The fragmentation potential $V_R(\eta ,T)$ at any temperature T, in
Eq.(\ref{eq:3}), is calculated within the Strutinsky renormalization procedure,
as
\bea
V_R(\eta ,T)&=&\sum_{i=1}^{2}{\Bigl [ V_{LDM}(A_i,Z_i,T)\Bigr ]}+
\sum_{i=1}^{2}{\Bigl [ \delta U_i\Bigr ]}exp(-\frac{{T}^2}{{T_0}^2})
\nonumber\\
&+&E_c(T)+V_P(T)+V_{\ell}(T),
\label{eq:10}
\eea
where the T-dependent liquid drop energy $V_{LDM}(T)$ is that of Ref.
\cite{davidson94} with the (Seeger's) constants at T=0 re-fitted to give the
experimental binding energies $B$ \cite{audi95}, defined as
$B=V_{LDM}(T=0)+\delta U$. The shell corrections $\delta U$ are calculated in
the "empirical method" of Myers and Swiatecki \cite{myers66}. Some of these
details are given in Appendix I. Figure 2 illustrates the kind of
comparisons obtained for $V(\eta )$ calculated at $R=C_1+C_2=C_t$ and
T=0 for the experimental and newly fitted binding energies. Apparently, the
binding energies fit within 1 to 1.5 MeV.

The $V_P$ is an additional attraction due to the nuclear proximity potential
\cite{blocki77}, which is also considered temperature-dependent here,
\be
V_P(R,T)=4\pi\bar{R}(T)\gamma b(T)\Phi (s,T),
\label{eq:41}
\ee
where $\bar{R}(T)$ and $\Phi (s,T)$ are, respectively, the inverse of the
root mean square radius of the Gaussian curvature and the universal function
which is independent of the geometery of the system, given by
\be
\Phi (s,T)=\left \{
\ba{ll}
-{1 \over 2}(s-2.54)^2-0.0852(s-2.54)^3 & \hbox{for} \quad s\le 1.2511 \\
-3.437exp(-{s \over 0.75}) & \hbox{for} \quad s\ge 1.2511
\ea
\right.
\label{eq:42}
\ee
\be
\bar{R}(T)=\frac{C_1(T)C_2(T)}{C_t(T)},
\label{eq:43}
\ee
and $\gamma$ is the specific nuclear surface tension given by
\be
\gamma =0.9517\left[1-1.7826\left(\frac{N-Z}{A} \right)^{2}
\right] MeV fm^{-2}.
\label{eq:44}
\ee
In Eq. (\ref{eq:42}), $s(T)$ (=${{R-C_t(T)}\over b(T)}$) is the overlap
distance, in units of b, between the colliding surfaces. The temperature
dependence in radii $R_i$ is given as \cite{davidson94,royer92},
\be
R_i(T)=r_0(T) A_i^{1\over 3}=1.07(1+0.01T) A_i^{1\over 3}
\label{eq:11}
\ee
with the surface width
\be
b(T)=0.99(1+0.009T^2).
\label{eq:12}
\ee
The same temperature dependence of R(T) is also used for Coulomb potential
$E_c(T)=Z_1Z_2e^2/R(T)$, where the charges $Z_i$ are fixed by minimizing the
potential $V_R(\eta ,T)$ in the charge asymmetry coordinate
$\eta _Z={{(Z_1-Z_2)}/{(Z_1+Z_2)}}$. The shell corrections $\delta U$ in
Eq. (\ref{eq:10}) are considered to vanish exponentially for $T_0=1.5$ MeV
\cite{jensen73}.

Also, for the angular momentum effects (so far included here for the calculation 
of total kinetic energy only)
\be
V_{\ell}(T)={{\hbar ^2\ell (\ell +1)}\over {2{I(T)}}}.
\label{eq:13}
\ee
In the non-sticking limit, where
$R_a=C_1(T)+C_2(T)+\Delta R=C_t(T)+\Delta R$,
the moment of inertia in (\ref{eq:13}) is given by
\be
I(T)={I}_{NS}(T)=\mu R_a^2.
\label{eq:14}
\ee
In this case, the separation distance $\Delta R$ is assumed to be beyond the 
range of nuclear proximity forces, which is about 2 fm . However, when 
$\Delta R$ is within the range of nuclear proximity ($<2$ fm), we get in the 
complete sticking limit
\be
I(T)={I}_{S}(T)=\mu R_a^2+{2\over 5}A_1mC_1^2+{2\over 5}A_2mC_2^2.
\label{eq:15}
\ee
For the $\ell$-value, in terms of the bombarding energy $E_{cm}$ of the
entrance channel $\eta _{in}$, we have
\be
\ell =\ell_{c}=R_a\sqrt{2\mu [E_{cm}-V(R_a,\eta _{in},\ell =0)]}/\hbar,
\label{eq:16}
\ee
or, alternatively, it could be fixed for the vanishing of fusion barrier.
In this work, however, we use $\ell =0$ for the IMF cross sections and take 
$\ell_{c}$ as a variable parameter for total kinetic energy (TKE) calculations 
(see Fig. 8).

The mass parameters $B_{\eta \eta}(\eta )$, representing the kinetic energy
part in Eq. (\ref{eq:3}), are the smooth classical hydrodynamical masses
\cite{kroeger80}, since we are dealing here with a situation where the
shell effects are almost completely washed out.

The assault frequency $\nu _0$, in Eq.(\ref{eq:2}), is given simply as
\be
\nu _0={{(2E_2/\mu )^{1/2}}\over R_0},
\label{eq:17} 
\ee
with the kinetic energy of the lighter fragment $E_2=(A_1/A) Q_{eff}$, for the
$Q_{eff}$ shared between the two fragments as inverse of their masses.
However, for the calculations of s-wave cross sections, instead of $\nu _0$, 
we use an empirically determined normalization constant.

Finally, the temperature dependent scattering potential V(R,T), normalised to
the exit channel binding energy, is
\be
V(R,T)=Z_1Z_2e^2/R(T)+V_P(T)+V_{\ell}(T). 
\label{eq:18}
\ee
This means that all energies are measured w.r.t $B_1(T)+B_2(T)$, and the
fragments go to ground state ($T\rightarrow 0$) via the emission of light
particle(s) and/or $\gamma$-rays of energy $E_x$.

\section{CALCULATIONS}
The calculations are made in two steps: (i) with temperature effects included
only in the shell corrections, i.e., using $\delta U(T)$, but T-independent
$V_{LDM}$ and V(R); and (ii) with temperature effects included also in both
the liquid drop energy and scattering potential, i.e., using $V_{LDM}(T)$,
$\delta U(T)$ and V(R,T). This allows us to study explicitly the role of
temperature in different terms of the potential. In both sets of the
calculations, we first take $\ell$=0, i.e., use $V_{\ell}$=0 through out, but
then study the effect of adding this term to the potential V(R) for
calculating the total kinetic energy (TKE) alone.

\subsection{Temperature effects only in shell corrections}
Figure 3 gives our calculated fragmentation potentials $V(\eta ,T)$ for
$^{56}Ni^*$ at T=0, as well as at other two temperatures refering to the
compound nucleus excitation energies $E_{CN}^*$ of the experiments of Ref.
\cite{sanders87,sanders89}. The R-values chosen are: $R=R_a=C_t$ at $T$=0, and, as
before \cite{gupta02}, $R=R_a+\overline{\Delta R}$ with
$\overline{\Delta R}=$0.30 and 0.31 fm, arbitrarily, for $T=$3.39 and 3.60 MeV
which correspond to the experimental energies $E_{cm}=$51.6 and 60.5 MeV,
respectively. The near independence of the structure in $V(\eta )$ on R-value
was studied in our earlier works \cite{gupta99a,gupta99b}. The
$\delta U$ at these temperatures reduce almost to zero. However, we notice
that the N=Z, A=4n $\alpha$-nucleus structure is obtained at all temperatures,
which has its origin apparently in the macroscopic liquid drop energy and is
due to the "Wigner term" in it, as was also shown earlier in Refs.
\cite{sharma00,gupta02}. Note that here the $V_{LDM}$ and other R-dependent
terms ($E_c$ and $V_P$) are not yet T-dependent (see next subsection). This
means that for use of only $\delta U$ as T-dependent, the N=Z $\alpha$-nuclei
fragments should be produced preferentially in the decay of $^{56}Ni^*$ at all
temperatures.

The preformation probability $P_0$ of the fragments, calculated for the
potentials in Fig. 3, is given in Fig. 4. The case of $T=0$ is not shown here
since cold $^{56}Ni$ (in the ground-state) can not decay because of its
negative Q-value. Interesting enough, for both the temperatures (the two
temperatures are nearly the same), the yields are large for only a small
window of $A\le 16$ fragments, including the light particles ($A\le 3$). Also,
the $\alpha$-nucleus fragments $^4He$, $^8Be$, $^{12}C$ and $^{16}O$ and the
light particle $^1H$ (in addition to the evaporation residues, not
included here) are preferentially preformed. This means that, out of all the
fragments observed in the decay of $^{56}Ni^*$, the ones with $A\le 16$ are
strongly preformed. The other ones with $A>16$, if observed, must have larger
penetrability P, since the decay constant is a combined effect of both the
preformation factor $P_0$ and penetrability P ($\nu _0$ is nearly constant).

Figure 5 gives the results of our calculation for the normalized decay
constants, equivalently, the s-wave production cross sections for {\it only} 
the most favoured (largest yields or cross sections) $\alpha$-nucleus 
fragments, compared with the experimental data at two energies, taken from Fig. 9 
of Ref. \cite{sanders89}. In the lower panel, the calculation at $E_{cm}$=51.6 MeV, 
using $\overline{\Delta R}$=0.3 fm, is fully normalized to the experimental data 
for the favoured $\alpha$-nucleus fragments only. Then, in the upper panel, for 
the higher energy $E_{cm}$=60.5 MeV we find that, for the use of the same 
normalization as obtained in lower panel and for a further normalization of the 
A=12 fragment yield, the best fit to the $\alpha$-nuclei fragment data is obtained 
for $\overline{\Delta R}$=0.29 fm, a value lower than that used for the lower 
incident energy $E_{cm}$=51.6 MeV. This is contrary to the expected behaviour of 
increased R at higher temeratures but, as we shall see below in Fig. 7, this is 
a result of our having not included here the contribution of angular momentum 
term in the fragmentation potential ($V_{\ell}=0$ in $V(\eta ,T)$) and hence in 
the cross sections. Also, the inclusion of temperature effects in other terms 
(the $V_{LDM}$, $E_c$ and $V_P$) are important, as is shown below in subsection 
III.B. Hence, Fig. 5 (and Fig. 7 below) show that the dynamical cluster-decay 
model contains the required structure of the measured yields (and TKEs) in this 
experiment \cite{sanders87,sanders89}.

Figure 6 shows the complete mass spectra for decay of $^{56}Ni^*$ calculated
at both the energies and compared with the measured yields \cite{sanders89}.  
The calculated yields are for the energetically favoured, most probable, mass 
fragments (see Figs. 3 and 4). Note that the experimental data in \cite{sanders87,sanders89}
are available only for fragments heavier than mass 11, and in steps of mass one 
for $E_{cm}$=60.5 MeV, but in steps of only mass two for $E_{cm}$=51.6 MeV due to
a deteriorated mass resolution at the lower bombarding energy. For
comparisons, the calculations are normalized to the experimental data for one 
fragment mass (A=20) only. The role of the penetrability P is evident in this 
figure, since some of the strongly preformed fragments, like $^{4}He$ and $^1H$ 
in Fig. 4, are now shown as less favoured decays (smaller cross sections, not 
shown in Fig. 6 since they lie below the chosen scale). The same is true of 
weakly preformed fragments (in Fig. 4), with $A>16$. Specifically, amongst the 
light particles, mass 3 fragment ($^3He$) is shown to be produced with a large 
cross section, and for lighter fragments ($A<12$), instead of A=8 ($^8Be$), the 
fragments with A=6 and 10 are shown to be produced with larger cross sections. 
This means that of all the residue products ($A\le 4$, not studied here) only 
mass 3 fragment ($^3He$) is produced and that the mass 4 ($^4He$) fragment is 
not at all produced as a dynamical cluster-decay fragment. This non-occurence 
of $^4He$ as a dynamical cluster-decay product, is an interesting result, giving 
a strong support to the credential of the model. For mass 8 ($Be$) decay, perhaps 
the contribution of higher $\ell$-values is important. For the heavier fragments
($A>20$), the calculated cross sections are rather small due to the fact that
here the contribution of only $\ell =0$ term is considered. Also, in experiments
it is difficult to separate the contributions of direct (such as alpha-transfer 
and orbiting processes) and compound nucleus yields for the heavy mass fragments (A$>20$) 
(see Ref. \cite{sanders99} and references therein). Thus, in view of the fact that 
we are dealing here with only the $\ell =0$ case and that the temperature effects 
are not included in full in the potential, the comparisons in Fig. 6 between the 
theory and experiments could be said atleast reasonable.

Figure 7 shows the results of our calculation for total kinetic energy (TKE),
with angular momentum $\ell$ effects included only in the scattering potential
V(R). We notice that the calculated TKEs for the sticking limit (using $I_S$) 
compare reasonably good with the experimental data. This means that, even though
$\overline{\Delta R}$ is non-zero (=0.29 and 0.3 fm), the sticking limit for
the moment of inertia is preferred. Also, unlike the
$\overline{\Delta R}$-values, the $\ell$-values required for the case of
higher energy data is now of a larger value ($\ell =25\hbar$ for $E_{cm}=$60.5
MeV as compared to $24\hbar$ for $E_{cm}=$51.6 MeV), as expected. The measured
TKEs are taken from Ref. \cite{sanders89}.

\subsection{Fully temperature-dependent potential}
Figure 8 shows the scattering potential $V(R,T,\ell)$ for temperature effects
included in all the terms of the potential (compare this figure for $\ell=0$,
with Fig. 1 where temperature effects are included in $\delta U$ only).
Notice that as $\ell$-value increases, the TKE($\overline{\Delta R}$)-value 
increases, since the decay path for all the $\ell$-values begins at $R=R_a$.
Figure 9 gives our calculated fragmentation potentials $V(\eta ,T)$. The
T-values chosen are the same as in Fig. 3, where temperature effects were
included only in the shell corrections. The R-values here are:
$R(T)=C_1(T)+C_2(T)+\overline{\Delta R}=C_t(T)+\overline{\Delta R}$, with
$\overline{\Delta R}$-values as shown in the figure. We notice in Fig. 9 that,
due to the inclusion of temperature effects in all terms, the minima in the
potential, which were earlier only for $\alpha$-nuclei, are now obtained for
both the $\alpha$ and non-$\alpha$ fragments. This happens, possibly, due to 
the pairing energy term $\delta (T)$ in formula (\ref{eq:A2}) of Davidson et al.
\cite{davidson94}, which goes to zero for $T>2$ MeV. Thus, with the addition of
temperature, not only the shell structure effects go to zero but also the
explicitly preferred $\alpha$-nucleus structure washes out. Also, we notice
that the light particles ($A\le 4$) structure changes; in particular, the
minimum at $^4He$ disappears and a shallow minimum at $^2H$ appears.

Figure 10 gives the preformation factors $P_0$ for the two experimentally
chosen temperatures only, since the ground-state (T=0) decay is not possible.
We notice that the formation yields are large only for light fragments
($A<16$) and are of the same orders as in Fig. 4, except that now the
non-$\alpha$ fragments are also preformed equally strongly. However, the
calculated decay constants, equivalently, the fragments 
(s-wave) 
production cross
sections, in Fig. 11 do not show much improvement in their comparisons with
experiments. The comparisons are now somewhat better for the heavier fragments
but the yields for fragments lighter than A=9 are very low, lying below the
chosen scale. On the other hand, the calculated TKEs in Fig. 12 compare nicely
(even better than in Fig. 7) with the experimental data. Only the case of
sticking limit is shown since the $\overline{\Delta R}$-values are still
within the proximity limits. Note that the $\ell$-dependent contribution is
so far added here only in the scattering potential $V(R,T)$ and not yet in
the fragmentation potential $V(\eta ,T)$ which is needed for both the
preformation factor and penetrability. This extention is being carried out.

\section{Summary}
In summary, we have reformulated for hot nuclear systems, the preformed
cluster-decay model (PCM) of Gupta and collaborators for ground-state decays 
and applied it for the first time to the decay of a light compound nucleus such 
as $^{56}Ni^*$ formed in the reaction $^{32}S+^{24}Mg$ carried out at two 
incident energies $E_{cm}$=51.6 and 60.5 MeV 
\cite{sanders87,sanders89}.
In this experiment, the mass spectra for fragments heavier than
mass 12 and the total kinetic energies (TKEs) for only the favoured
$\alpha$-nucleus fragments are measured. Also, at another energy, in between
the two above, an enhanced yield is observed for $^8Be$ over the two
$\alpha$-particles emission \cite{thummerer01}. Our calculations are made first 
for the temperature effects included only in shell corrections and then in all 
terms of the potential, and in each case for $\ell$=0 only. The contribution due 
to $\ell$ is added only for estimating the TKEs.
Similar to the saddle-point model \cite{sanders91} and/ or the scission-point
model \cite{matsuse97}, the deformations of the fragments are taken into
account by the parametrization of the neck-in zone, proposed by Gupta and
collaborators \cite{khosla90,gupta97,kumar97}. This quantity is $\eta$-dependent
and could be calculated but is taken as a parameter here, which is the only 
parameter of the model.

For the temperature effects included in shell corrections only, we find that
the $\alpha$-nucleus fragments are favourably preformed and are due to the
macroscopic liquid drop energy alone since the shell effects are almost zero
at the energies under consideration. The calculated decay constants or the
normalized 
s-wave 
cross sections, in particular for the $\alpha$-nucleus fragments
are found to contain the complete structure of the experiments for a nuclear
shape with fragments separated by about 0.3 fm which is within the limits of
nuclear proximity effects. Some of the light particles (other than the ones
constituting the evaporation residue, not included here) are also predicted to 
be there in the mass spectra, but $^4He$ is shown to be absent. With angular
moments effects included, the calculated TKEs are found to compare rather
nicely with experimental data for the moment of inertia calculated for a
sticking limit.

For the full temperature effects in the potential, the non-$\alpha$ fragments
are also preformed equally strongly as the $\alpha$-nucleus fragments. The
cluster decay process now occurs at a somewhat larger separation distance,
which is also temperature dependent. Hence, the TKEs for a sticking moment of
inertia are now in somewhat better agreement with the experiments. However,
the comparison between the calculated 
(s-wave) 
and measured mass spectra is not
improved much, which calls for the inclusion of $\ell$-dependent potential in
the calculations of yields also, which is underway.

\section{Appendix I: Temperature dependent binding energies}
In Eq. (\ref{eq:10}) we have defined, within the Strutinsky renormalization
procedure, the binding energy B of a nucleus at temperature T as the sum of
liquid drop energy $V_{LDM}(T)$ and shell correction $\delta U(T)$,
\be
B(T)=V_{LDM}(T)+\delta Uexp(-\frac{{T}^2}{{T_0}^2}).
\label{eq:A1}
\ee
The $T$ dependent liquid drop part of the binding energy $V_{LDM}(T)$ used
here is that of Davidson et al. \cite{davidson94}, based on the semi-empirical
mass formula of Seeger \cite{seeger61}, as
\bea
V_{LDM}(T)&=&\alpha (T)A+\beta (T)A^{2\over 3}+\Bigl (\gamma (T)-{\eta (T)\over {A^{1\over 3}}}\Bigr )
\nonumber\\
&&\times \Bigl ({{I^2+2\mid I\mid }\over A}\Bigr )+{Z^2\over {r_0(T)A^{1\over 3}}}\Bigl (1-
{0.7636\over {Z^{2\over 3}}}
\nonumber\\
&-&{2.29\over {[r_0(T)A^{1\over 3}]^2}}\Bigr )+\delta (T){f(Z,A)\over {A^{3\over 4}}},
\label{eq:A2}
\eea
where
$$I=a_{a}(Z-N),\quad\hbox{$a_{a}=$1,}$$
and, respectively, for even-even, even-odd, and odd-odd nuclei,
$$f(Z,A)=(-1,0,1).$$

For $T=$0, Seeger \cite{seeger61} obtained the constants, by fitting all
even-even nuclei and 488 odd-A nuclei available at that time, as
$$\alpha (0)=-16.11\hbox{MeV,}\quad\beta (0)= 20.21\hbox{MeV,}$$
$$\gamma (0)=20.65\hbox{MeV,}\quad\eta (0)=48.00\hbox{MeV,}$$
with the pairing energy term 
$$\delta (0)=33.0\hbox{MeV,}$$
from Ref. \cite{benedetti64}.
Evidently, these constants need be re-fitted since a large amount of data has
become available \cite{audi95}, particularly for neutron-rich nuclei.
We found that the measured binding energies could be fitted within 1 to
1.5 MeV by changing the bulk constant $\alpha (0)$ and introducing a proton,
neutron asymmetry constant $a_a$. The $\alpha (0)$ works as an overall scaling
factor and $a_a$ controls the curvature of the experimental parabola (and
hence helps to fit the binding energies for neutron-rich nuclei), as expected.
Table 1 gives the new $\alpha (0)$ and $a_a$ constants for all the known
nuclei with $1\le Z\le 28$, relevant to present problem. The kind of
comparisons obtained between the experimental and calculated binding energies
is already illustrated in Fig. 2.

The T-dependent constants in Eq. (\ref{eq:A2}) were obtained numerically by
Davidson et al. \cite{davidson94} for the available experimental information
on excited states of 313 nuclei in the mass region $22\le A\le 250$ by
determining the partition function ${\cal{Z}}$(T) of each nucleus in the
canonical ensemble and making a least squares fit of the excitation energy
$$E_{ex}(T)=V_{LDM}(T)-V_{LDM}(T=0)$$
to the ensemble average 
$$E_{ex}(T)=T^2{\partial\over {\partial T}}ln{\cal{Z}}(T).$$
The $\alpha (T)$, $\beta (T)$, $\gamma (T)$, $\eta (T)$ and $\delta (T)$ thus
obtained are given in Figure 1 of Ref. \cite{davidson94} for $T\le 4$MeV,
extrapolated linearly for higher temperatures. For the bulk constant
$\alpha (T)$, instead, an empirically fitted expression to a Fermi gas model
is used, as
$$\alpha (T)=\alpha (0)+{T^2\over 15}.$$
Also, the $\delta (T)$ is constrained to be positive definite at all
temperatures, with $\delta (T>2 MeV)=$0. Finally, the analytical form for
$r_0(T)$, taken from Ref. \cite{brack74}, is
$$r_0(T)=1.07(1+0.01T).$$

For the shell corrections $\delta U$ in Eq. (\ref{eq:A1}), since there is no
microscopic shell model known that gives the shell corrections for light
nuclei, we use the empirical formula of Myers and Swiatecki \cite{myers66}.
For spherical shapes,
\be
\delta U=C\left [{{F(N)+F(Z)}\over {({A/2})^{2\over 3}}}-cA^{1\over 3}\right ]
\label{eq:A3}
\ee
where 
\be
F(X)={3\over 5} \left ( {{M_i^{5\over 3}-M_{i-1}^{5\over 3}} \over {M_i-M_{i-1}}}\right )
\left (X-M_{i-1}\right )-{3\over 5}\left (X^{5\over 3}-M_{i-1}^{5\over 3}\right )
\label{eq:A4}
\ee
with X=N or Z, $M_{i-1}<X<M_i$ and $M_i$ as the magic numbers 2,8,14 (or 20),
28,50,82,126 and 184 for both neutrons and protons. The constants C=5.8 MeV and
c=0.26. In this paper, we refer to the use of magic numbers 14 or 20 as MS14
or MS20 parametrization.

\section{ACKNOWLEDGMENTS}
This work is supported in parts by the Council of Scientific and Industrial
Research (CSIR), India, the VW-Stiftung, Germany, and the ULP/IN2P3, France.


\newpage
\par\noindent
{\bf FIGURE CAPTIONS}
\begin{description}
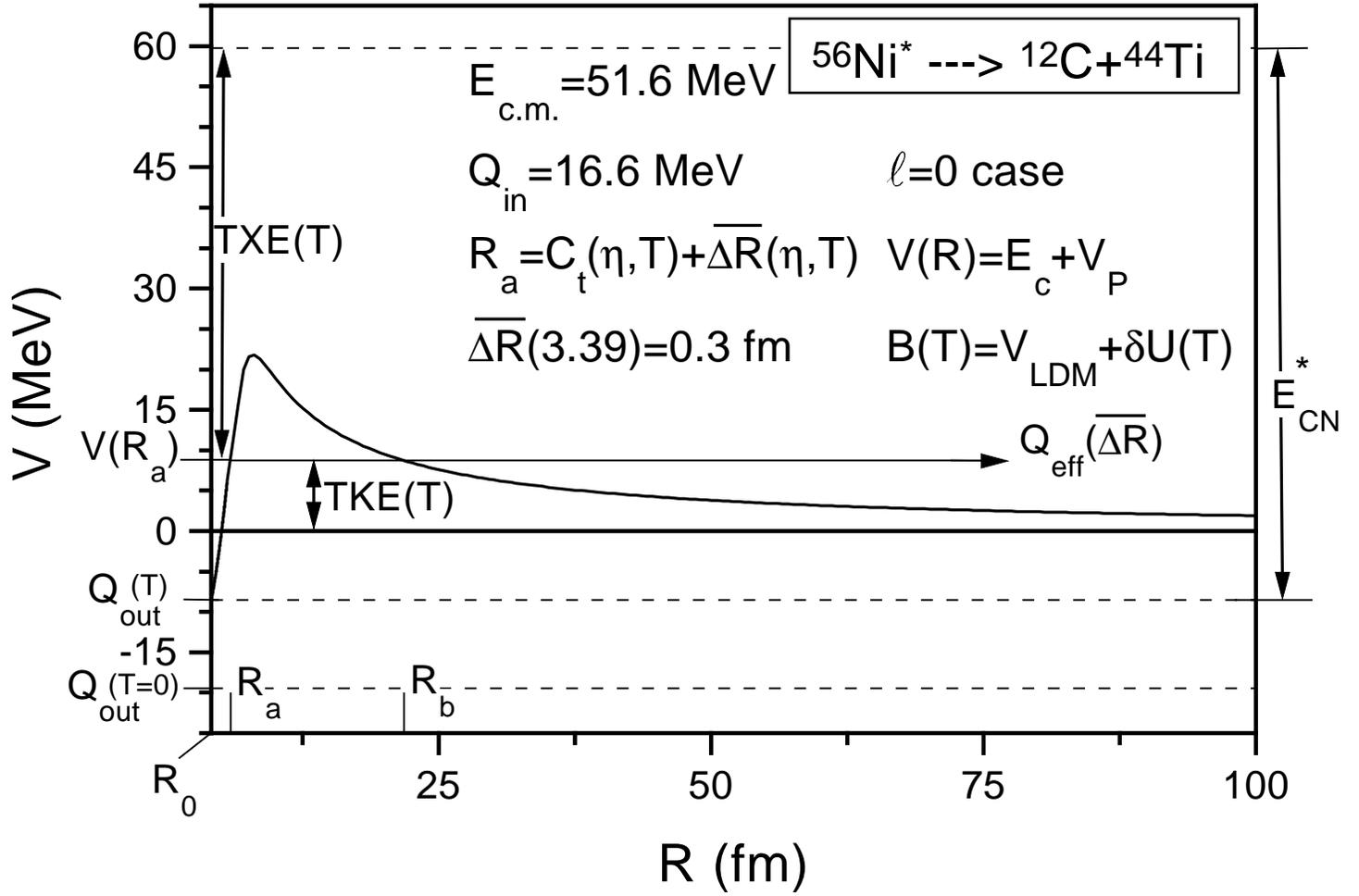
\item{Fig.1} The s-wave ($\ell =0$) scattering potential for
$^{56}Ni^*\rightarrow ^{12}C+^{44}Ti$, calculated for {\bf no} temperature
effects in $E_c$ and $V_P$, i.e. $V(R)=E_c+V_P$. The Q-values are calculated
from T-dependent binding energy $B(T)=V_{LDM}+\delta U(T)$. The actually
calculated decay path for
$V(R_a)=Q_{eff}(\overline{\Delta R})=V(C_t(T)+\overline{\Delta R})$ is shown,
where $\overline{\Delta R}$ is an average of the separation distances for 
different fragmentations (different $\eta$-values). 
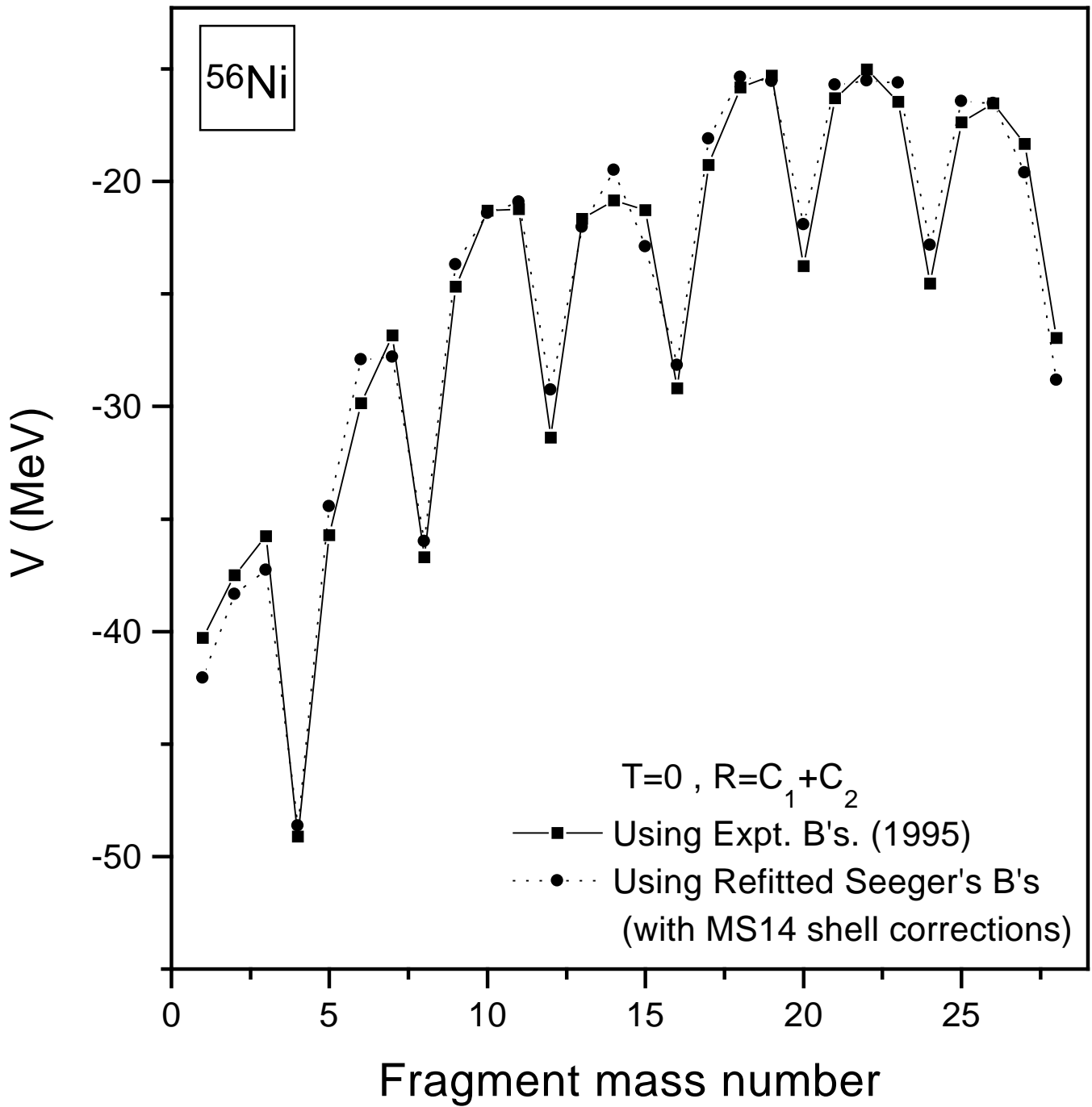
\item{Fig.2} The fragmentation potential for $^{56}Ni$ at T=0, $R=C_t$, using
the experimental binding energies (solid squares) \cite{audi95} and the
empirically fitted Seeger's binding energies (solid circles) with the new
constants of Table 1. Here, MS14 means the shell corrections from the
empirical method of Myers and Swiatecki \cite{myers66} with Z and N=14 as the
magic numbers.
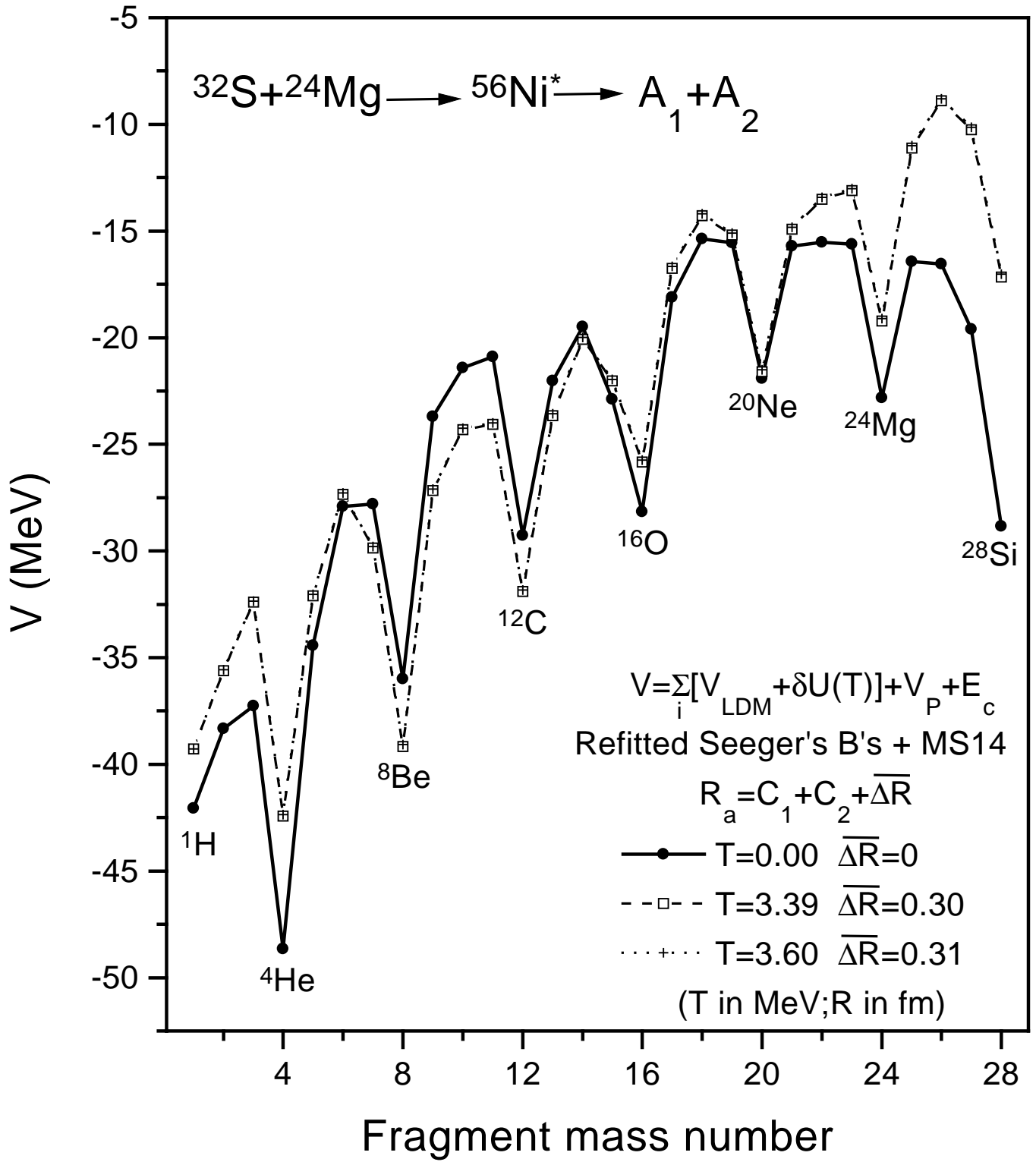
\item{Fig.3} The fragmentation potentials $V(\eta ,R,T)$ for $^{56}Ni^*$
compound system, calculated at the ground-state ($T$=0, $R_a=C_t$)
and at various temperatures with $R_a=C_1(T)+C_2(T)+\overline{\Delta R}$
values as shown. The $T$-dependence is included only in the shell corrections.
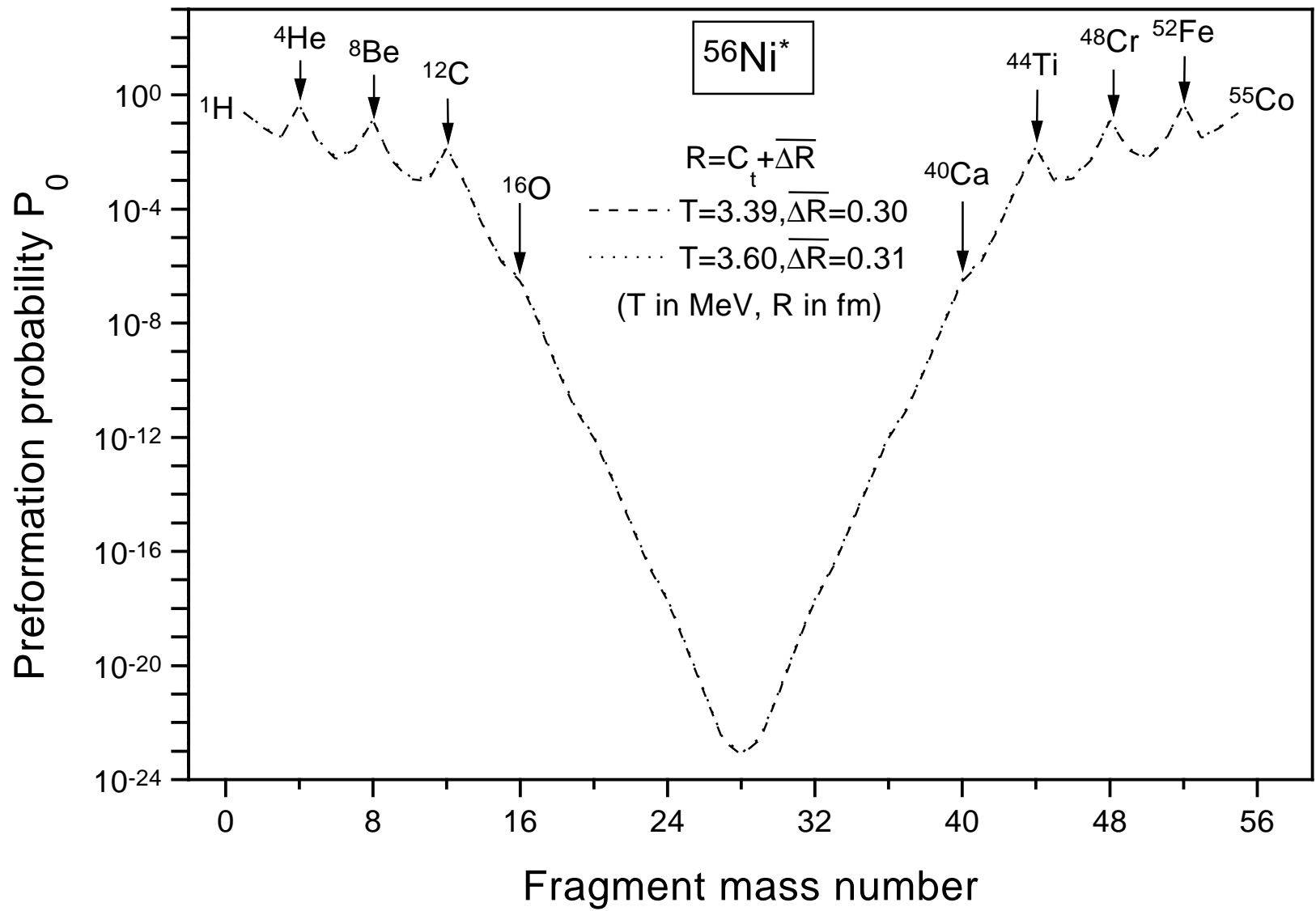
\item{Fig.4} The fragment preformation probability $P_0$ for $^{56}Ni^*$,
calculated by using the fragmentation potentials in Fig. 3 for the two
experimental T-values only.
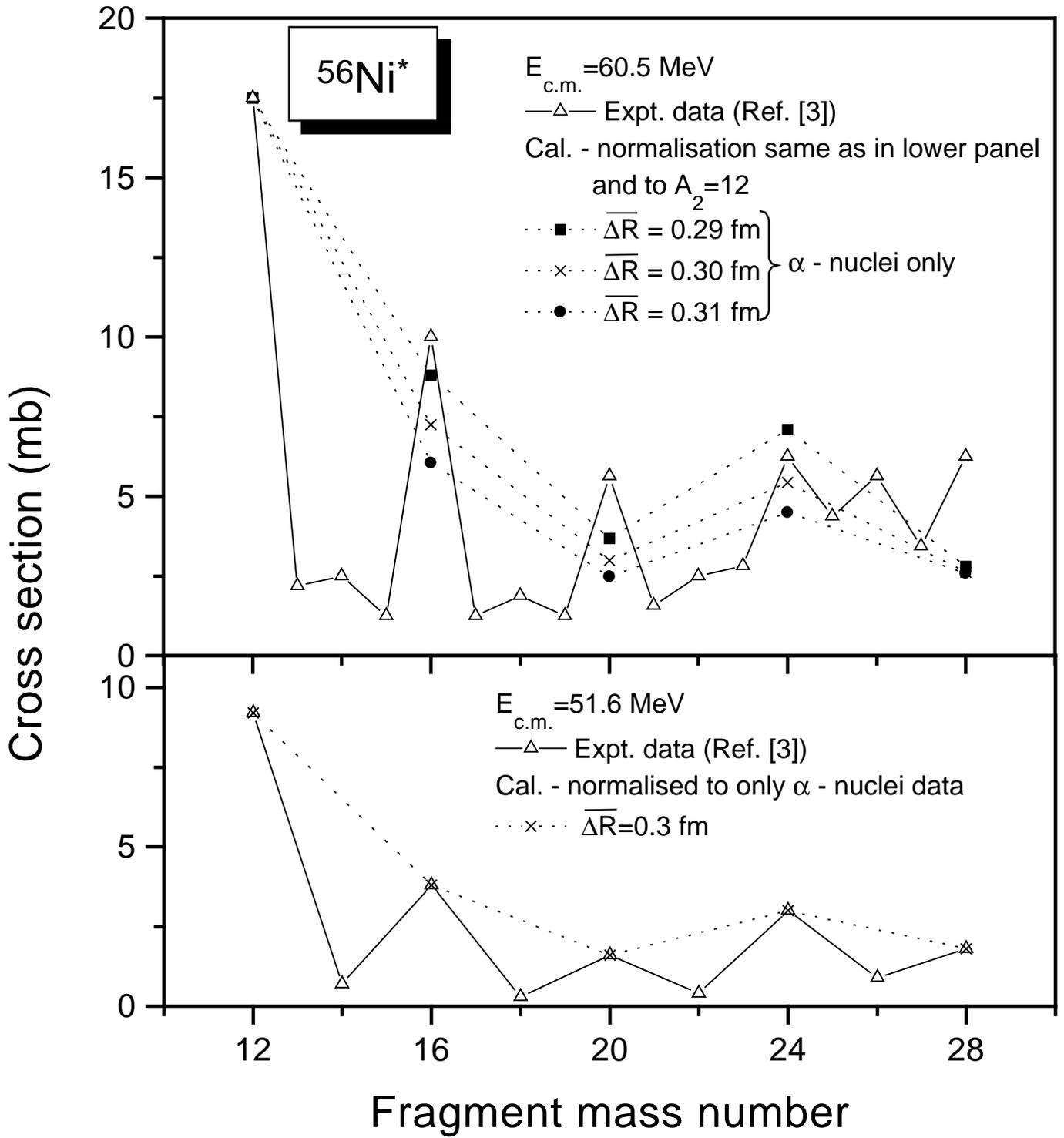
\item{Fig.5} The calculated s-wave cross sections for the $\alpha$-nucleus 
fragments compared with the measured ones produced in the reaction
$^{32}S+^{24}Mg\rightarrow ^{56}Ni^*$ at $E_{cm}=$51.6 and 60.5 MeV. The data 
are from Fig. 9 of Ref. \cite{sanders89}. The calculations for $E_{cm}=$51.6 
MeV in the lower panel are made for $\overline{\Delta R}=$0.30 fm and are 
normalized completely to the experimental data. Using the same normalization, 
the calculations for $E_{cm}=$60.5 MeV in the upper panel are made for 
$\overline{\Delta R}=$0.29, 0.30 and 0.31 fm and compared with the experimental 
data, for a further normalization of the data at fragment mass A=12. Only the
$\alpha$-nucleus fragments are studied, since they have the largest cross 
sections. The dotted lines are drawn only for the guide of eyes.
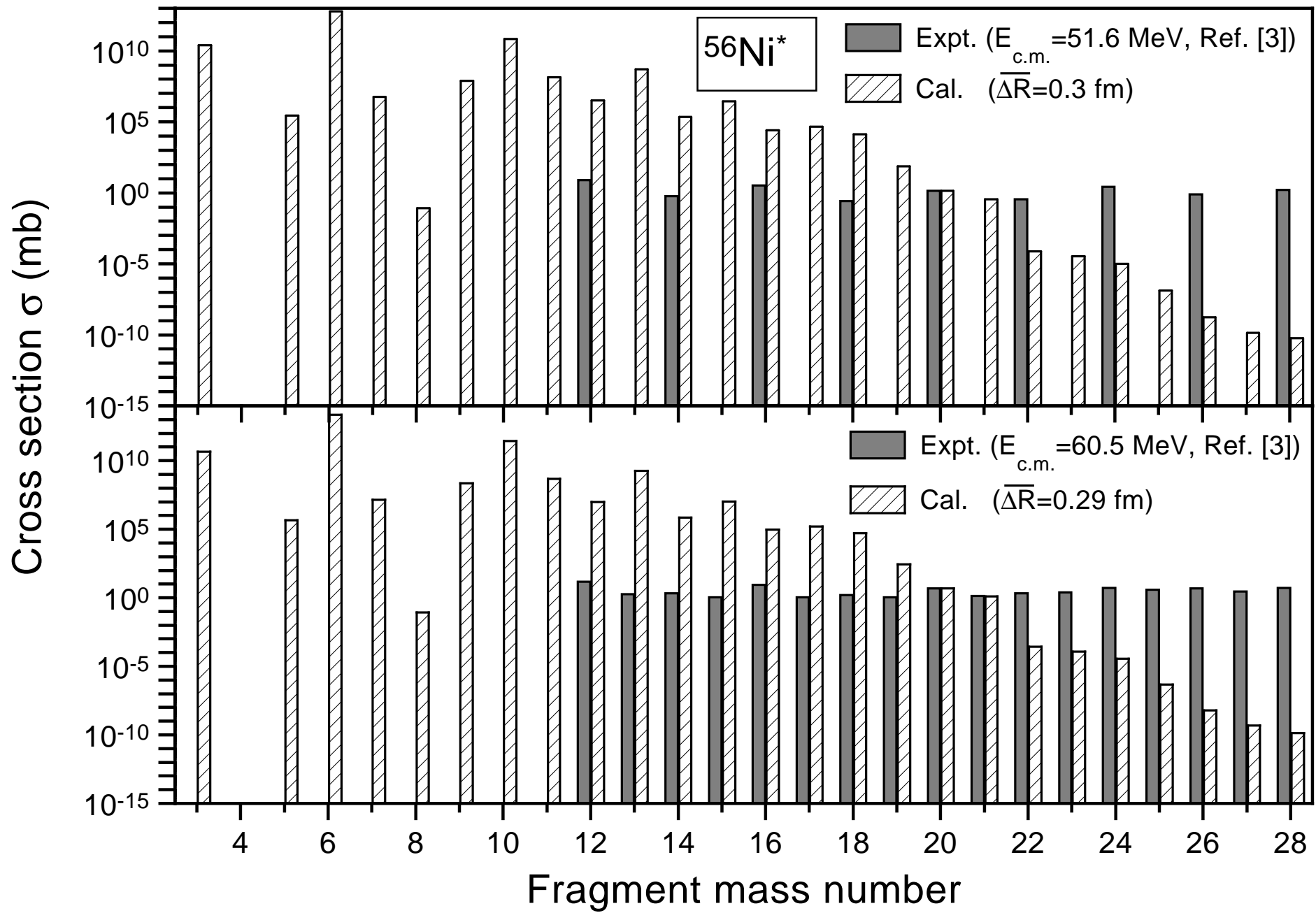
\item{Fig.6} Same as for Fig. 5, but studied for all the fragments at 
$E_{cm}=$51.6 MeV, $\overline{\Delta R}=$0.30 fm (upper panel) and 
$E_{cm}=$60.5 MeV and $\overline{\Delta R}=$0.29 fm (lower panel). 
The calculations are normalized to the experimental data for one fragment 
mass (A=20) only. The calculated (s-wave) cross sections are for the 
energetically most favoured fragments in $\eta$ coordinate i.e. fragments lying 
at the minimum in the fragmentation potential $V(\eta)$, minimized in $\eta_Z$ 
coordinate.
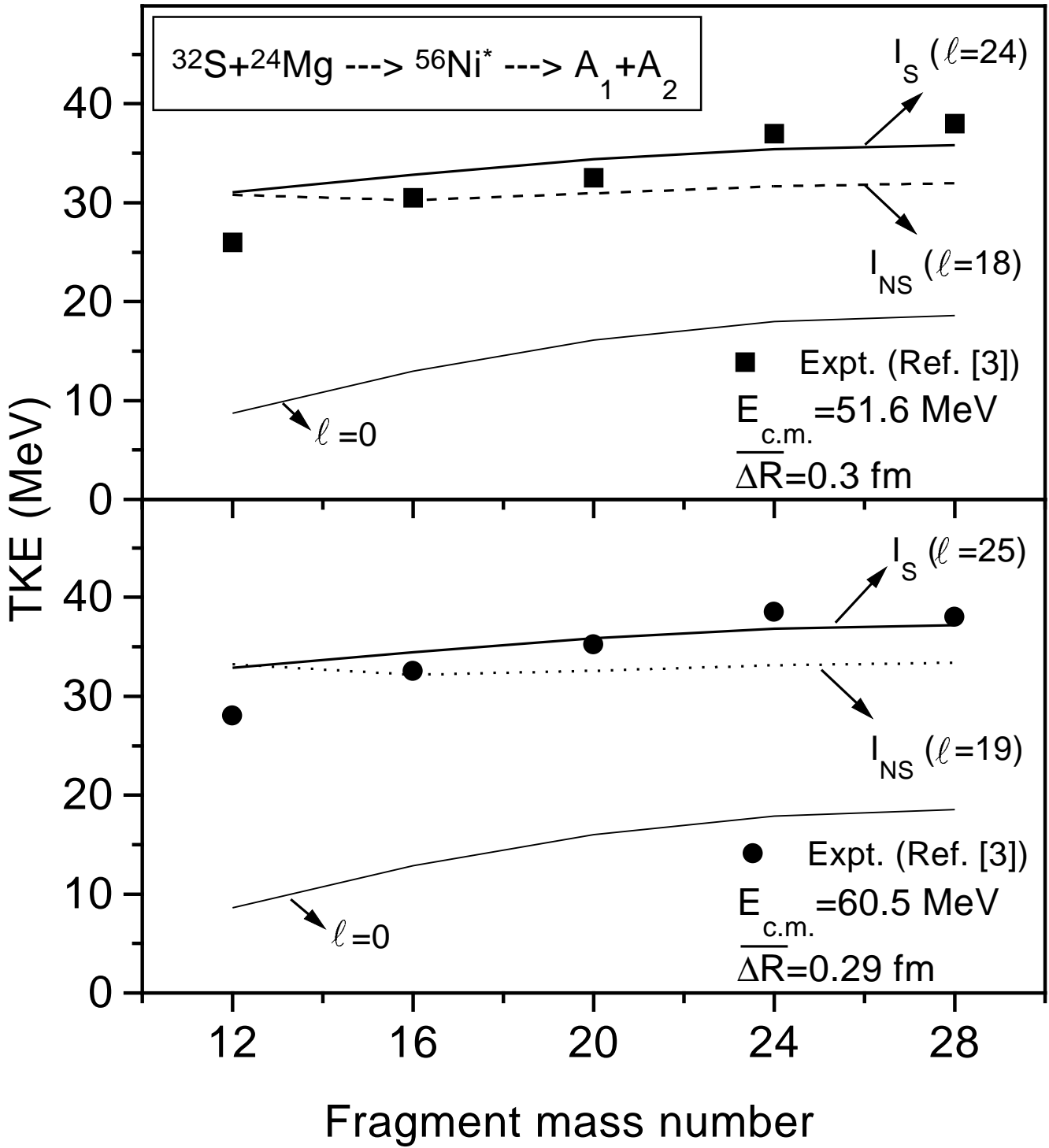
\item{Fig.7} The measured and calculated total kinetic energy (TKE) for
average $\overline{\Delta R}$ for the the reaction
$^{32}S+^{24}Mg\rightarrow ^{56}Ni^*\rightarrow A_1+A_2$, at the
two incident energies. The calculations for $\ell \ne 0$ are made for both the
cases of sticking and non-sticking limits (see text). The data are from Fig. 5 
(summed over all the angles) of Ref. \cite{sanders89}. The same data are also 
given in Fig. 10 of Ref. \cite{sanders91}, where it should be noted that Fig. 10(a)
refers to $E_{c.m.}$=60.5 MeV and Fig. 10(b) to $E_{c.m.}$=51.6 MeV.
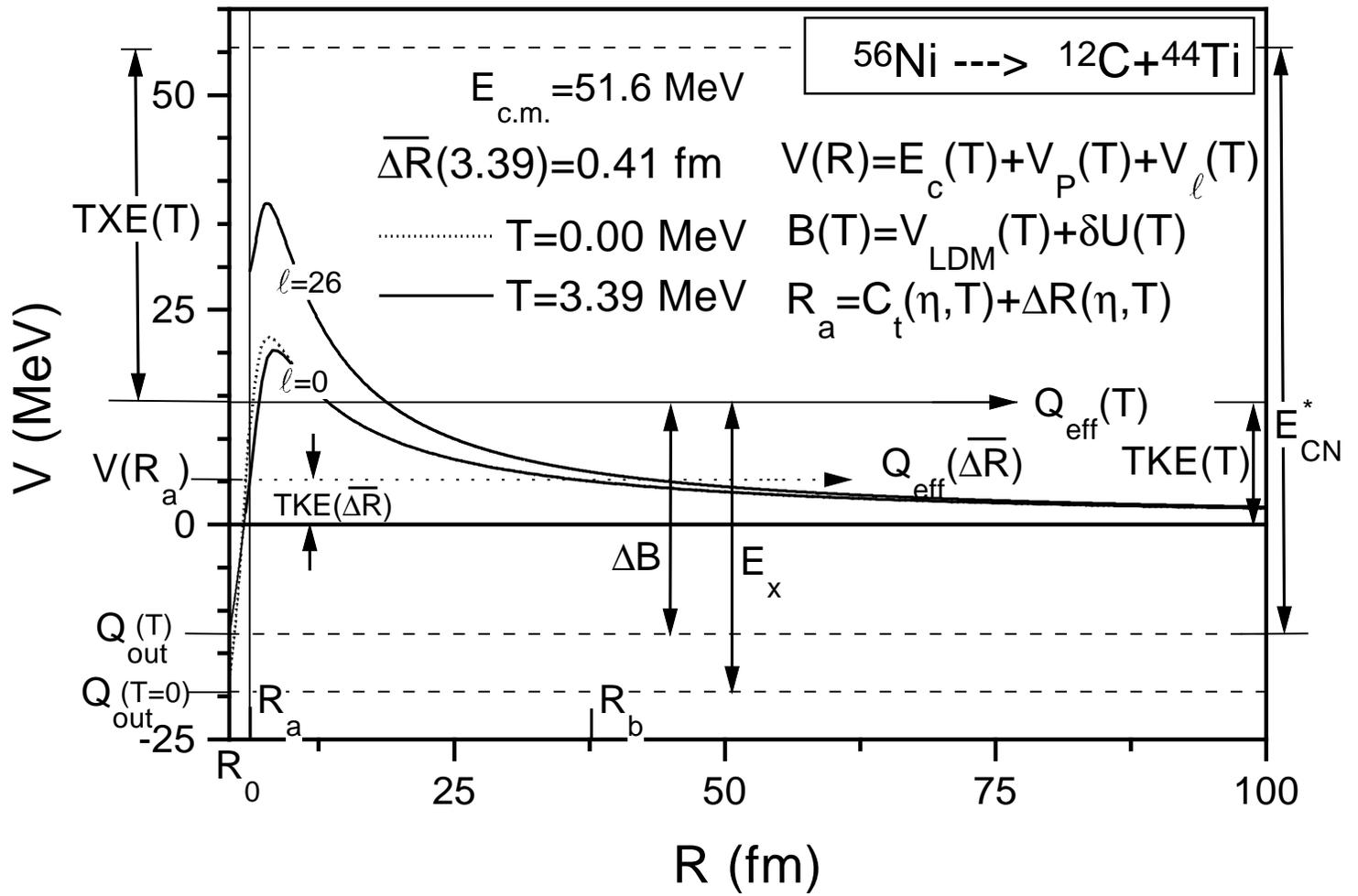
\item{Fig.8} Same as for Fig. 1, but with $\ell$, and T dependences included in 
$E_c$ and $V_P$ also, i.e., the scattering potential is
$V(R,T,\ell)=E_c(T)+V_P(T)+V_{\ell}(T)$ with Q-value now calculated from
$B(T)=V_{LDM}(T)+\delta U(T)$. Only the sticking limit of moment of inertia is
used in $V_{\ell}(T)$. The T=0 potential is shown for comparisons.
For all $\ell$-values, the decay path (dotted line), shown for $\overline{\Delta R}$, 
begins at $R=R_a$ (marked explicitly). The distribution of energies and definitions 
of other quantities like $\Delta B$ and $E_x$ are indicated for the calculated 
$\Delta R$-value. 
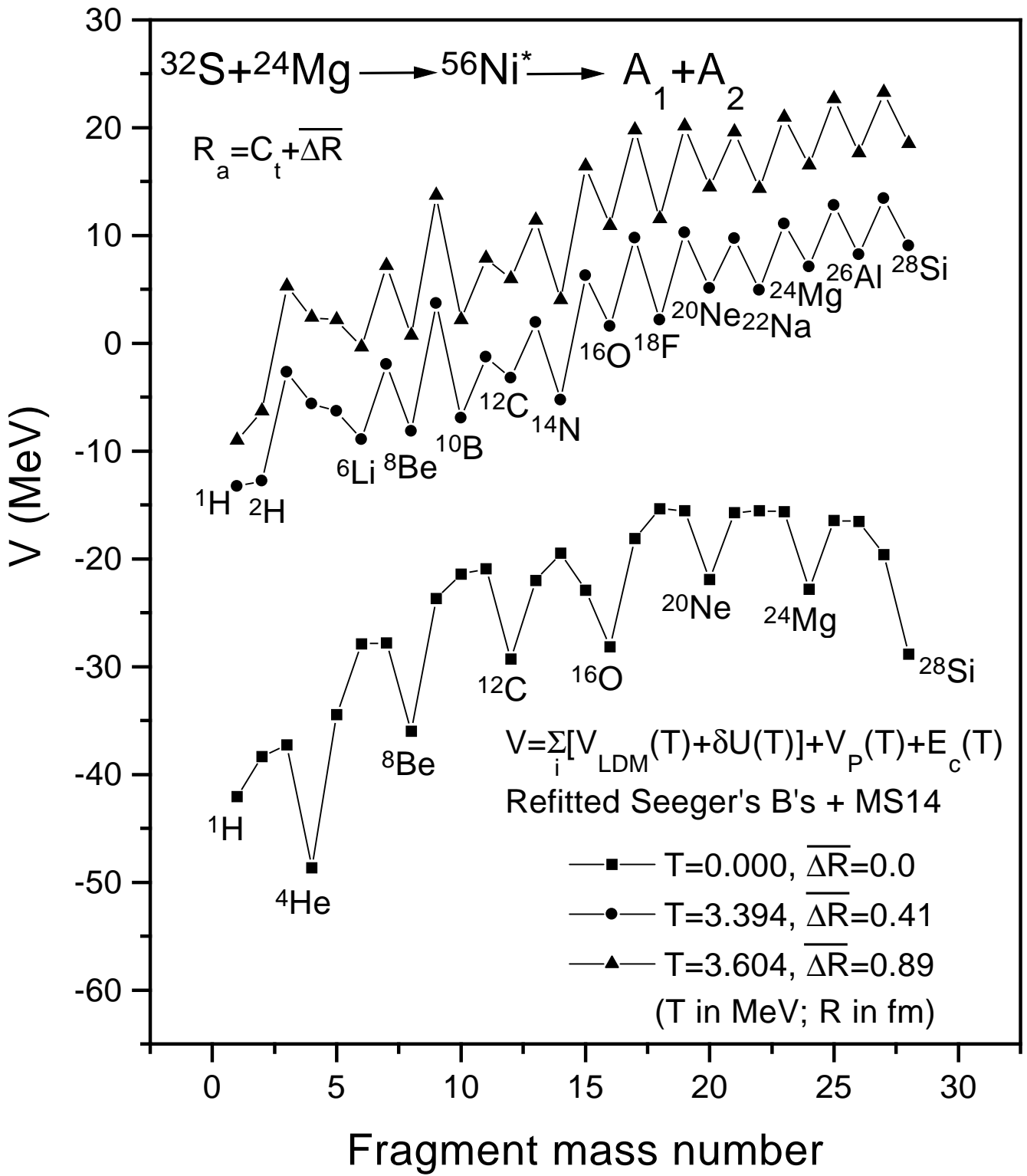
\item{Fig.9} Same as for Fig. 3, but for T-dependence in all the terms of the
fragmentation potential, and at $\overline{\Delta R}$ values as shown.
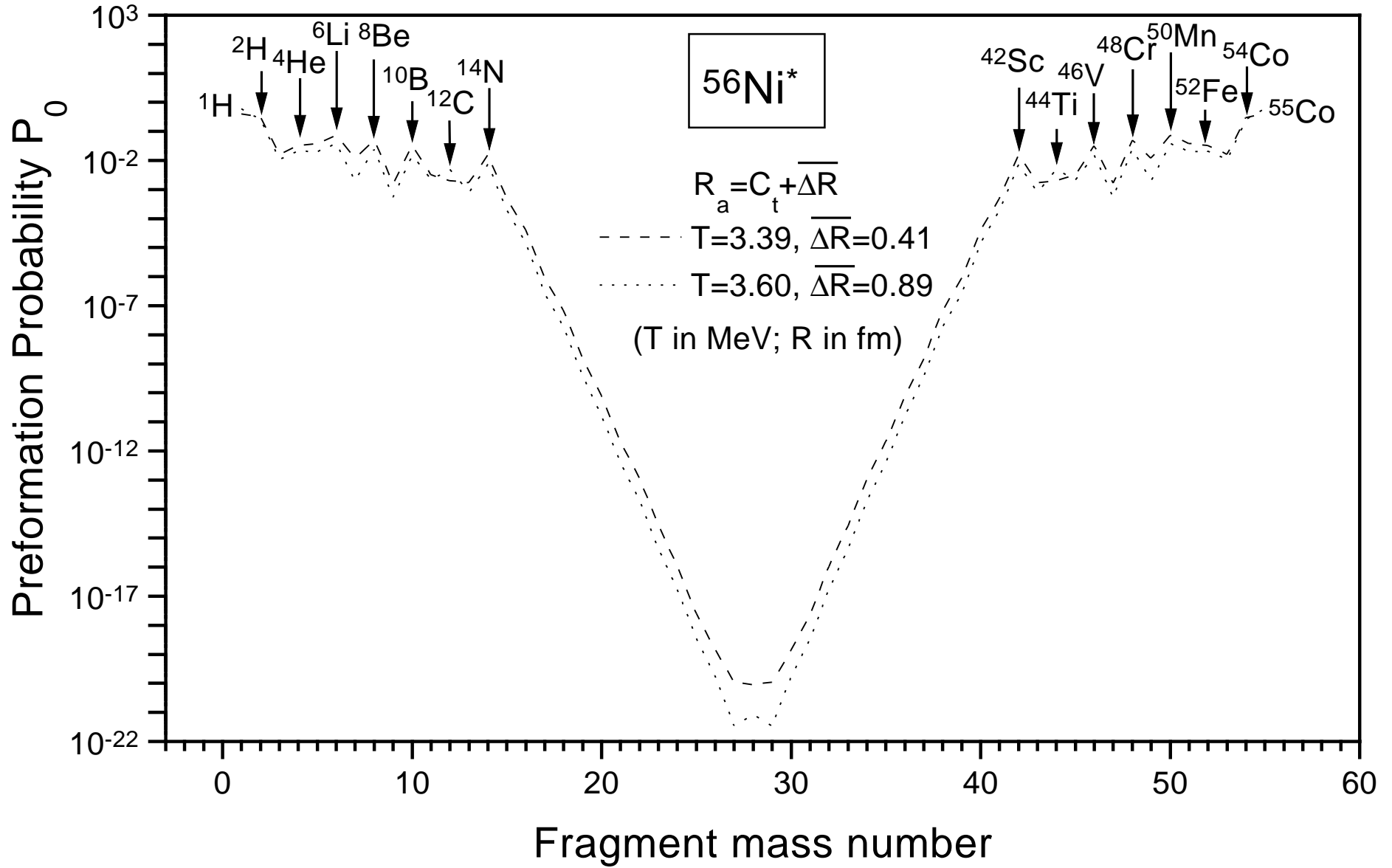
\item{Fig.10} Same as for Fig. 4, but for the fragmentation potential of
Fig. 9.
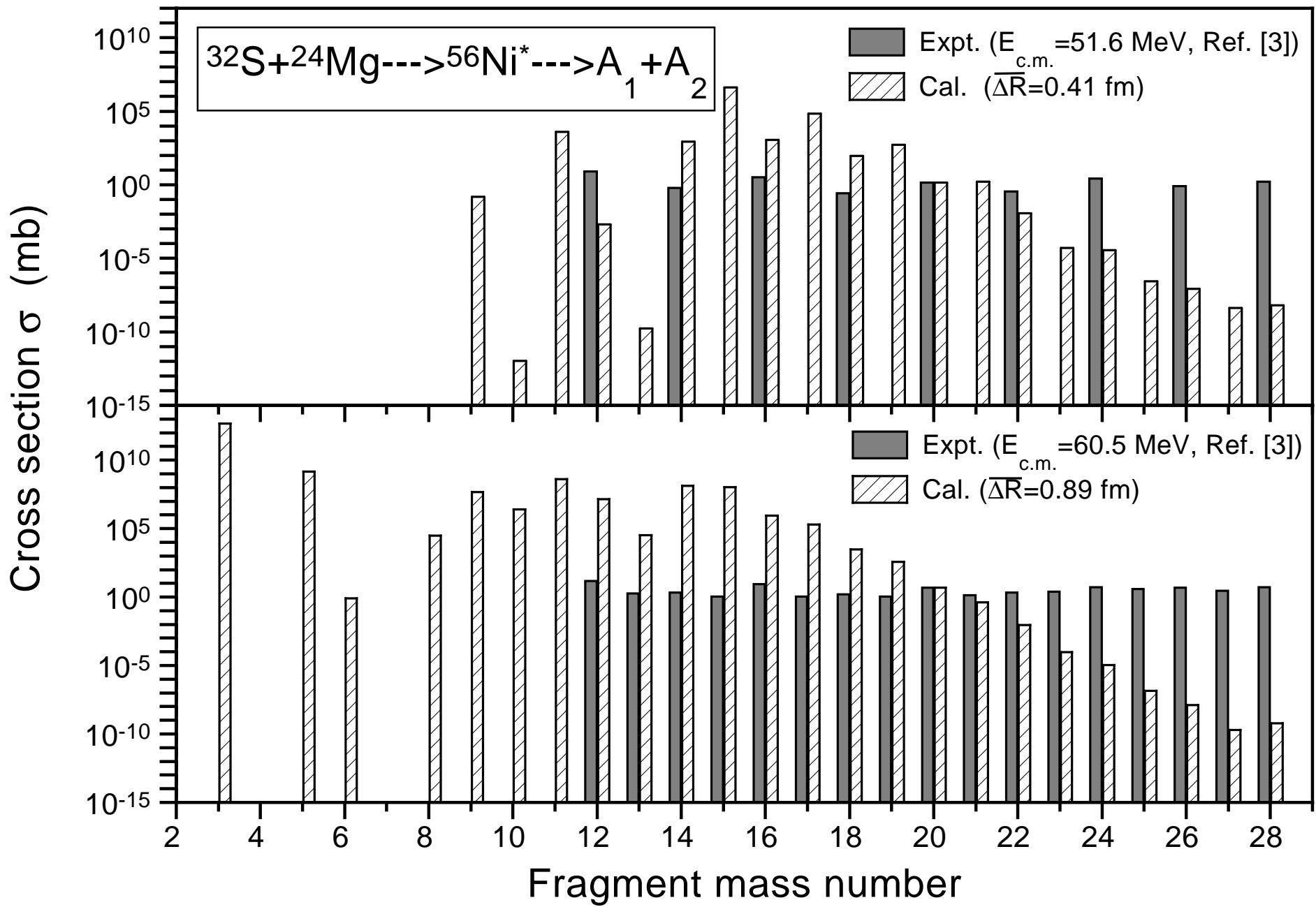
\item{Fig.11} Same as for Fig. 6, but for T-dependence in all the terms of the
fragmentation potential, and at $\overline{\Delta R}$ values as shown.
For lighter fragments, the calculated yields are not shown as they lie below
the chosen scale.
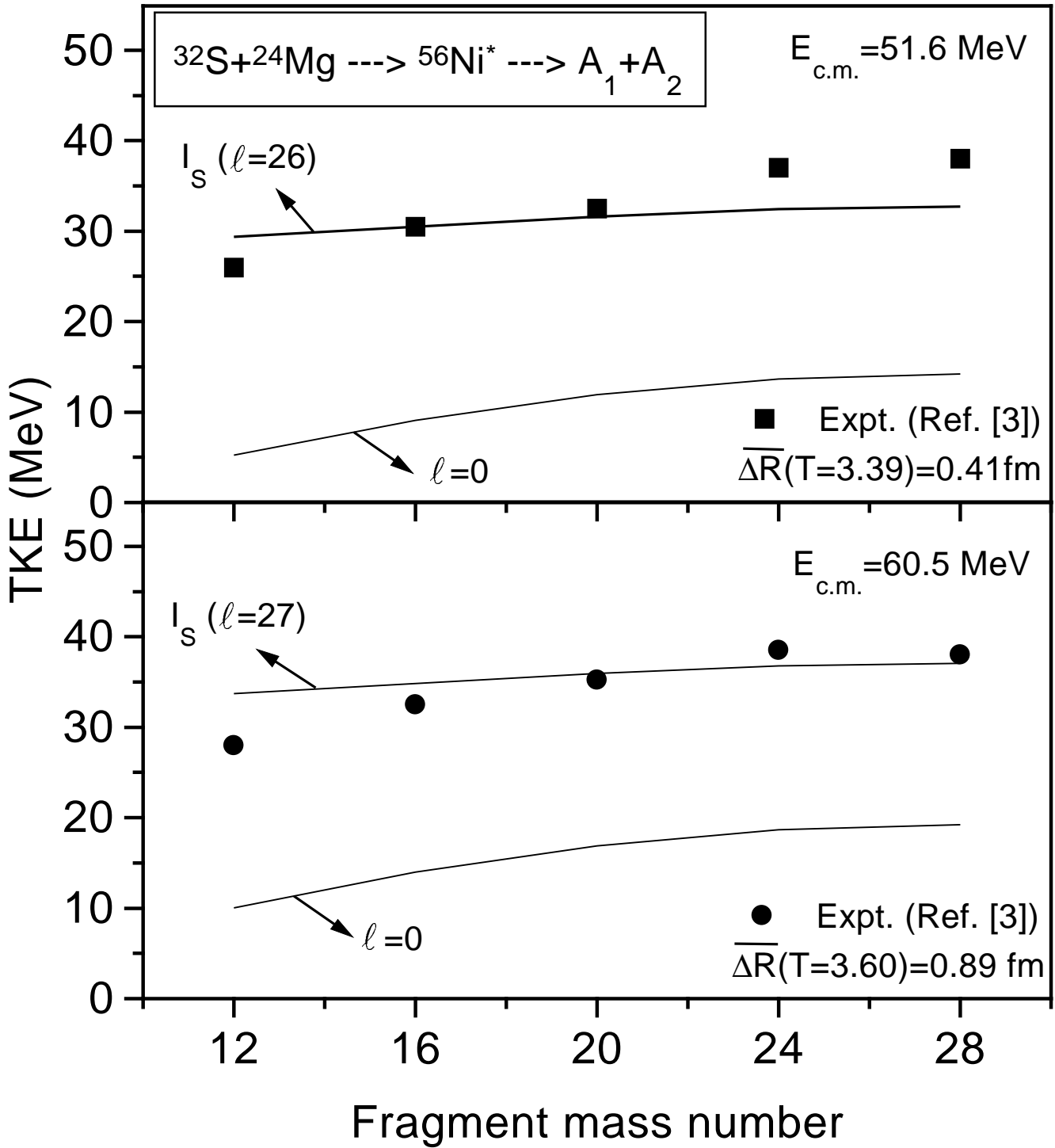
\item{Fig.12} Same as for Fig. 7, but for T-dependence in all the terms of the
fragmentation potential, and at $\overline{\Delta R}$ values as shown.
\end{description}

\newpage
\pagestyle{empty}
\addtolength{\oddsidemargin}{0.0cm}
\addtolength{\evensidemargin}{0.0cm}
\begin{table}{\bf Table 1.} Re-fitted bulk and asymmetry constants for Seeger's mass formula.\\[1ex]
\begin{tabular}{|c|c|c|c||c|c|c|c||c|c|c|c|}\hline
Z & N & $\alpha (0)$ & a$_a$ & Z & N & $\alpha (0)$ & a$_a$ & Z & N & $\alpha (0)$ & a$_a$ \\ \hline

1&  2  &-15.85&0.10& 6&    9         &-15.70&0.10&10&      7          &-15.70&0.50 \\ \hline
 &  3  &-16.95&0.12&  &   10         &-15.10&0.10&  &      8          &-15.90&0.90 \\ \hline
 &  4  &-13.00&0.05&  &   11         &-14.80&0.10&  &     13          &-15.95&0.50 \\ \hline
 &  5  &-13.70&0.12&  &12,13,15,16   &-15.00&0.80&  &     14          &-15.70&0.50 \\ \hline
2&  1  &-15.50&0.10&  &   14         &-14.85&0.80&  & 9-12,15-22      &-16.16&0.88 \\ \hline
 &  2  &-16.00&0.10& 7&    3         &-14.30&0.20&11&      7          &-15.55&0.50 \\ \hline
 &  3  &-16.80&0.30&  &    4         &-15.20&0.50&  &      8          &-15.80&0.50 \\ \hline
 & 4,5 &-14.20&0.30&  &    5         &-16.20&0.80&  &     14          &-15.95&0.50 \\ \hline
 &  6  &-13.50&0.10&  &    6         &-16.55&0.80&  & 9-13,15-24      &-16.20&0.86 \\ \hline
 & 7,8 &-13.00&0.10&  &    7         &-16.80&0.80&12&   8-10          &-16.11&0.90 \\ \hline
3&1,2,4,5&-16.60&0.10&  &    8         &-16.30&0.80&  &   11-25         &-16.20&0.86 \\ \hline
 &  3  &-16.98&0.98&  &    9         &-16.20&0.80&13&   8-10          &-16.11&0.90 \\ \hline
 &  6  &-13.80&0.98&  &  10,11       &-15.90&0.94&  &   11-26         &-16.22&0.84 \\ \hline
 &  7  &-14.30&0.40&  &   12         &-15.75&0.94&14&   8-12          &-16.11&0.90 \\ \hline
 & 8,9 &-13.20&0.10&  &   13         &-15.80&0.94&  & 13-20,27,28     &-16.28&0.84 \\ \hline
4&  1  &-13.00&0.01&  &   14         &-15.65&0.94&  &    21-26        &-16.22&0.84 \\ \hline
 &  2  &-14.50&0.10&  &   15         &-15.90&0.94&15& 9-13,20-31      &-16.30&0.82 \\ \hline
 &  3  &-16.20&0.80&  &   16         &-16.00&0.94&  &    14-19        &-16.36&0.78 \\ \hline
 &  4  &-16.98&0.98&  &   17         &-16.10&0.93&16& 10-14,21-28     &-16.30&0.82 \\ \hline
 &  5  &-16.70&0.60& 8&    4         &-14.00&0.94&  &    15-20        &-16.40&0.78 \\ \hline
 &  6  &-15.50&0.80&  &    5         &-15.25&0.94&  &    29-33        &-16.32&0.80 \\ \hline
 &  7  &-15.30&0.50&  &    6         &-15.90&0.94&17&11-14,20,21,29-34&-16.36&0.78 \\ \hline
 &  8  &-14.30&0.10&  &    7         &-16.35&0.94&  &    15-19        &-16.45&0.78 \\ \hline
 &  9  &-14.00&0.10&  &    8         &-16.20&0.94&  &    22-28        &-16.32&0.82 \\ \hline
 & 10  &-13.30&0.01&  &    9         &-16.18&0.94&18&12-14,21,22,31-35&-16.36&0.78 \\ \hline
5&  2  &-14.60&0.10&  &   10         &-15.95&0.94&  &    15-20        &-16.45&0.78 \\ \hline
 &  3  &-16.50&0.10&  &   11         &-15.93&0.94&  &    23-30        &-16.32&0.78 \\ \hline
 &  4  &-16.60&0.60&  & 12,14        &-15.85&0.94&19&13,14,22,23,30-36&-16.38&0.78 \\ \hline
 &  5  &-16.99&0.10&  &   13         &-15.90&0.94&  &    15-21        &-16.44&0.78 \\ \hline
 &  6  &-16.60&0.60&  &   15         &-16.10&0.94&  &    24-29        &-16.36&0.80 \\ \hline
 &  7  &-16.30&0.10&  &   16         &-16.15&0.90&20& 14,15,22-37     &-16.38&0.78 \\ \hline
 &  8  &-15.35&0.10&  &   17         &-16.30&0.92&  &    16-21        &-16.48&0.78 \\ \hline
 &  9  &-15.10&0.10&  &   18         &-16.11&0.92&21& 15-23,31-38     &-16.42&0.77 \\ \hline
 & 10  &-14.45&0.10& 9&    5         &-15.25&0.90&  &    24-30        &-16.38&0.78 \\ \hline
 & 11  &-14.10&0.10&  &    6         &-15.90&0.90&22&    16-39        &-16.42&0.77 \\ \hline
 & 12  &-13.45&0.10&  &    7         &-16.28&0.90&23&    17-40        &-16.42&0.77 \\ \hline
 & 13  &-13.10&0.10&  &    9         &-16.30&0.90&24&    18-25        &-16.45&0.77 \\ \hline
 & 14  &-13.00&0.40&  &   10         &-16.15&0.90&  &    26-41        &-16.42&0.77 \\ \hline
6&  2  &-13.00&0.10&  &8,11,17,19,20&-16.20&0.90&25&    19-26        &-16.46&0.77 \\ \hline
 &  3  &-13.85&0.80&  &   12         &-16.01&0.90&  &    27-42        &-16.42&0.77 \\ \hline
 &  4  &-15.70&0.10&  &   13         &-16.05&0.90&26&    19-43        &-16.46&0.77 \\ \hline
 & 5,7 &-16.50&0.10&  &   14         &-15.95&0.90&27&    21-28        &-16.48&0.77 \\ \hline
 &  6  &-16.65&0.10&  &15,16,18      &-16.11&0.90&  &    29-45        &-16.46&0.77 \\ \hline
 &  8  &-15.90&0.10&10&    6         &-15.25&0.50&28&    22-48        &-16.48&0.77 \\ \hline
\end{tabular}                                                    
\end{table}
\end{document}